%% file: main.tex
\documentclass[11pt]{article}

\usepackage[a4paper,margin=2.4cm]{geometry}
\usepackage[T1]{fontenc}
\usepackage[utf8]{inputenc}
\usepackage{lmodern}
\usepackage{microtype}
\usepackage{amsmath}
\usepackage{amssymb}
\usepackage{bm}
\usepackage{graphicx}
\usepackage{booktabs}
\usepackage{array}
\usepackage{longtable}
\usepackage{xcolor}
\usepackage{hyperref}
\newcolumntype{L}[1]{>{\raggedright\arraybackslash}p{#1}}

\hypersetup{
  colorlinks=true,
  linkcolor=blue!45!black,
  citecolor=blue!45!black,
  urlcolor=blue!45!black
}

\newcommand{\CPtwoK}{\textsc{CP2K}}
\newcommand{\Quickstep}{\textsc{Quickstep}}
\newcommand{\ii}{\mathrm{i}}
\newcommand{\dd}{\mathrm{d}}
\newcommand{\BZ}{\mathrm{BZ}}
\newcommand{\bR}{\bm{R}}
\newcommand{\bk}{\bm{k}}
\newcommand{\bq}{\bm{q}}
\newcommand{\br}{\bm{r}}
\newcommand{\ba}{\bm{a}}
\newcommand{\bb}{\bm{b}}
\newcommand{\bG}{\bm{G}}

\newcommand{\Tr}{\mathrm{Tr}}
\numberwithin{equation}{section}
\hypersetup{pdfauthor={Johann V. Pototschnig; Jan Kloppenburg; Shaojie He; Chengze Zhou; Shiyang Wang; Sam Azadi; Hossein Mirhosseini; Juerg Hutter; Thomas D. Kuehne}}
\title{Symmetry-reduced k-point sampling and Wannier analysis with localized Gaussian orbitals}

\author{Johann V. Pototschnig$^{a,b}$, Jan Kloppenburg$^{a,b}$, Shaojie He$^{c}$,\\
Chengze Zhou$^{d}$, Shiyang Wang$^{e}$, Sam Azadi$^{f,g}$, Hossein Mirhosseini$^{a,b}$,\\
J\"urg Hutter$^{h}$, and Thomas D. K\"uhne$^{a,b,i,*}$\\[0.6em]
\small $^{a}$Center for Advanced Systems Understanding (CASUS), Untermarkt 20,\\
\small D-02826 G\"orlitz, Germany\\
\small $^{b}$Helmholtz-Zentrum Dresden-Rossendorf (HZDR), Bautzner Landstrasse 400,\\
\small D-01328 Dresden, Germany\\
\small $^{c}$ School of Computer Science and Technology,
Harbin University of Science and Technology,\\ 
\small 150001 Harbin City, China\\
\small $^{d}$Department of Chemistry, School of Science, Tianjin University, 300072 Tianjin, China\\
\small $^{e}$ Dalian Institute of Chemical Physics, Chinese Academy of Sciences, 116023 Dalian, China\\
\small $^{f}$Department of Physics and Astronomy, University of Manchester,\\
\small Oxford Road, Manchester M13 9PL, United Kingdom\\
\small $^{g}$Department of Physics, Clarendon Laboratory, University of Oxford,\\
\small Parks Road, Oxford OX1 3PU, United Kingdom\\
\small $^{h}$Department of Chemistry, University of Zurich, Winterthurerstrasse 190,\\
\small CH-8057 Zurich, Switzerland\\
\small $^{i}$Chair of Computational Systems Science, Institute of Artificial Intelligence,\\
\small Technische Universit\"at Dresden, Helmholtzstrasse, D-01069 Dresden, Germany\\
[0.4em]
\small $^{*}$Corresponding author: \href{mailto:tkuehne@cp2k.org}{tkuehne@cp2k.org}}
\date{}

\begin{document}

\maketitle

\begin{abstract}
Brillouin-zone sampling with localized orbitals requires consistent
transformations between sparse real-space matrices and complex Bloch
states. Symmetry reduces the number of reciprocal-space problems, but
scalar integration weights alone cannot reconstruct the matrices and
wavefunctions needed for densities and orbital analysis.
We develop a unified formulation of these transformations for
nonorthogonal Gaussian bases and describe its implementation in \CPtwoK.
Atom permutations, local orbital rotations, and lattice phases connect
irreducible-point solutions to full symmetry orbits while preserving
the overlap metric. Direct and fast lattice transforms couple this
construction to self-consistent calculations, including complex orbital
optimization and finite-temperature occupations. Gauge-consistent
full-mesh orbitals and directed inter-point overlaps support Wannier90
export, Wannier construction, and Hamiltonian interpolation.
In silicon and aluminium tests, symmetry reduces a 64-point mesh to four
representatives while retaining cell energies or electronic free energies
within $1.5\times10^{-12}$ hartree of the unreduced calculation.
Hybrid exchange and many-body applications illustrate the additional
role of auxiliary bases and transfer momenta. The resulting framework
connects symmetry-reduced integration to gauge-consistent analysis of
periodic electronic states within a localized-orbital representation.
\end{abstract}


\clearpage
\section{Introduction}
\label{sec:introduction}

Periodic electronic-structure calculations replace an extended material
by a finite cell with Bloch boundary conditions~\cite{Bloch1929}. The
remaining integration over the Brillouin zone determines how reliably
that cell represents the bulk. Large insulating supercells often require
only the zone center, whereas compact cells and metallic occupations
demand explicit sampling. Converging the self-consistent-field (SCF) equations at a
single k-point does not remove this quadrature error. Conversely, reducing
a mesh by symmetry should preserve its physical content, not change the
approximation being solved.

Localized atomic orbitals offer an efficient representation for
density-functional theory (DFT) in the Kohn--Sham (KS)
formulation~\cite{Hohenberg1964,Kohn1965}. Periodic localized-basis methods
are well established, including Gaussian-orbital and numerical
atom-centered implementations~\cite{Dovesi2014,Blum2009}. Their locality
makes real-space matrices sparse, but their nonorthogonality introduces
a k-dependent overlap metric. The present work addresses the
transformations needed to preserve both properties when exploiting
crystal symmetry and evaluating quantities that connect different
k-points.

We use \CPtwoK{} as the implementation platform. Its \Quickstep{} module
combines the Gaussian and plane wave (GPW) method with the Gaussian
augmented plane wave (GAPW) all-electron
extension~\cite{Lippert1997,Lippert1999,VandeVondele2005a,
Kuehne2020CP2K,Iannuzzi2026CP2K}. Orbitals are expanded in localized
Gaussian atomic orbitals (AOs), while charge densities and electrostatic
potentials use multigrids and plane waves. Atom-pair blocks are stored
in the distributed block-compressed sparse row (DBCSR)
representation~\cite{Borstnik2014}. Each block carries an image-cell
index as well as basis-function indices. Bloch matrices are assembled
from these blocks by phase-weighted lattice sums, and the density matrix
is returned to the same image-cell representation after solving the
reciprocal-space problem. The use of these methods in \emph{ab initio}
molecular dynamics is reviewed by Hutter, Iannuzzi, and
K\"uhne~\cite{HutterIannuzziKuehne2024}.

In the current work we extended symmetry reduction to this construction.
A scalar observable that is invariant on a k-point orbit can be evaluated using the orbit's
representative and multiplicity. Reconstructing a matrix or wavefunction
also requires atom permutations, angular-momentum
rotations, and image-cell phases. These operations must preserve the
overlap metric and be consistent with the inverse lattice transform.
Neglecting this distinction can leave band energies apparently
reasonable while corrupting the reconstructed density or exported
orbitals.

Wannier construction provides a complementary requirement: a full-mesh
neighbor graph and a consistent orbital gauge. Directed inter-point
overlaps retain phase and metric information that scalar Brillouin-zone
averages do not contain.

A unified, implementation-level
formulation linking symmetry-covariant reconstruction of
nonorthogonal matrix orbits, recovery of full-mesh Bloch orbitals, and
directed inter-k-point overlaps for Wannier analysis
is presented in this work. 
The underlying Bloch and Wannier theories are established.
The contribution lies in their consistent realization through sparse
Gaussian image-cell operators and explicit symmetry mappings.
Silicon and aluminium calculations test equivalence at fixed quadrature.
The formulation is then connected to orbital optimization, finite-temperature
occupations, and the distinct momentum structure of hybrid and many-body
methods.

Section~\ref{sec:quadrature} introduces reciprocal-space quadrature,
and Sections~\ref{sec:bloch}--\ref{sec:k290-spglib} establish the basis
and symmetry conventions. Sections~\ref{sec:transforms}--\ref{sec:implementation}
describe lattice transforms, self-consistency, and implementation.
Sections~\ref{sec:analysis} and \ref{sec:advanced-kpoint-methods} develop
the applications, followed by numerical verification in
Section~\ref{sec:verification}. The Supporting Information records
source versions, implementation coverage, and reproduction details.

\section{Brillouin-zone quadrature and boundary conditions}
\label{sec:quadrature}

Consider a periodic one-particle quantity.  A bulk observable per primitive cell can be written as
a Brillouin-zone average,
\begin{equation}
  \langle O\rangle =
  \frac{\Omega_{\mathrm{cell}}}{(2\pi)^3}
  \sum_{n}
  \int_{\BZ}
  f_{n\bk}\,
  o_{n}(\bk)\,
  \dd\bk ,
  \label{eq:bz-average}
\end{equation}
where $\Omega_{\mathrm{cell}}$ is the real-space cell volume,
$f_{n\bk}$ is an occupation number, and $o_n(\bk)$ is the band-resolved
contribution.  The prefactor follows from
\begin{equation}
  \Omega_{\BZ}=\frac{(2\pi)^3}{\Omega_{\mathrm{cell}}},
\end{equation}
so that Eq.~\eqref{eq:bz-average} is equivalently an average over the
Brillouin-zone volume.  A k-point calculation replaces this integral by a
quadrature,
\begin{equation}
  \frac{\Omega_{\mathrm{cell}}}{(2\pi)^3}
  \int_{\BZ} F(\bk)\,\dd\bk
  \approx
  \sum_{i=1}^{N_k} w_i F(\bk_i),
  \qquad
  \sum_{i=1}^{N_k} w_i = 1.
  \label{eq:bz-quadrature}
\end{equation}
The quadrature must be consistent with the occupations, crystal symmetries,
and phase conventions used to construct the electronic density.

The Gamma-point approximation is the special quadrature with
\begin{equation}
  N_k=1,\qquad \bk_1=\bm{0},\qquad w_1=1.
\end{equation}
Its accuracy improves when the integrand varies weakly over the small
Brillouin zone of a large cell, but cell size alone is not a convergence test.
If a primitive cell is repeated
$N_1\times N_2\times N_3$ times, the reciprocal lattice vectors shrink to
\begin{equation}
  \bb_{\alpha}^{\mathrm{SC}}=\frac{1}{N_{\alpha}}\bb_{\alpha},
\end{equation}
and the Gamma point of the supercell folds the primitive-cell mesh
\begin{equation}
  \bk_{\bm{m}} =
  \frac{m_1}{N_1}\bb_1+
  \frac{m_2}{N_2}\bb_2+
  \frac{m_3}{N_3}\bb_3,
  \qquad
  m_\alpha=0,\ldots,N_\alpha-1,
  \label{eq:supercell-folding}
\end{equation}
into the supercell Brillouin-zone center. This exact folding equivalence
requires a translationally repeated structure and an electronic state with
the same primitive-cell periodicity. It does not turn a disordered or
symmetry-broken supercell into a primitive-cell calculation. Nor are the
equivalent representations computationally identical in a
localized-basis code. A supercell increases the number of atoms and basis
functions, whereas a k-point mesh increases the number of independent dense
eigenproblems built from the same local basis.

The need for k-points is especially transparent for the electronic density,
\begin{equation}
  \rho(\br)=
  \sum_{\sigma}
  \sum_{n}
  \int_{\BZ}
  \frac{\Omega_{\mathrm{cell}}\,\dd\bk}{(2\pi)^3}
  f_{n\bk}^{\sigma}
  \left|\psi_{n\bk}^{\sigma}(\br)\right|^2,
  \label{eq:density-integral}
\end{equation}
where the Bloch orbitals are normalized per cell, or, after quadrature,
\begin{equation}
  \rho(\br)\approx
  \sum_{\sigma}
  \sum_{i=1}^{N_k} w_i
  \sum_n f_{ni}^{\sigma}
  \left|\psi_{ni}^{\sigma}(\br)\right|^2 .
  \label{eq:density-quadrature}
\end{equation}
For an insulator the functions entering Eq.~\eqref{eq:density-integral} are
usually smooth in $\bk$, and moderate meshes converge rapidly.  For a metal,
the occupation contains a Fermi-surface discontinuity,
\begin{equation}
  f_{n\bk} =
  \Theta\!\left(\mu-\varepsilon_{n\bk}\right)
  \qquad (T_{\mathrm{el}}=0),
  \label{eq:metal-step}
\end{equation}
which makes the integral much harder.  Smearing replaces
Eq.~\eqref{eq:metal-step} by a smooth function, but the total number of
electrons must still be conserved by a global chemical potential,
\begin{equation}
  N_{\mathrm{el}} =
  \sum_{\sigma}
  \sum_{i=1}^{N_k} w_i
  \sum_n
  f\!\left(\varepsilon_{ni}^{\sigma}-\mu\right).
  \label{eq:global-electron-count-intro}
\end{equation}
This is why k-point sampling is tied directly to the occupation machinery and
cannot be treated as a purely local diagonalization option.

Bloch's theorem supplies the structure that makes the quadrature feasible.  A
crystalline orbital satisfies
\begin{equation}
  \psi_{n\bk}(\br+\bR)=
  \exp(\ii\bk\cdot\bR)\psi_{n\bk}(\br),
  \label{eq:bloch-boundary}
\end{equation}
or, equivalently,
\begin{equation}
  \psi_{n\bk}(\br)=
  \exp(\ii\bk\cdot\br)u_{n\bk}(\br),
  \qquad
  u_{n\bk}(\br+\bR)=u_{n\bk}(\br).
  \label{eq:bloch-periodic-part}
\end{equation}
Different k-points are therefore different boundary conditions on the same
periodic Hamiltonian.  In a localized Gaussian basis this statement becomes a
phase relation between basis functions in different image cells.  The
implementation described below is built around preserving that phase relation
exactly while reusing the sparse real-space matrices of the Gamma-point code.

\section{Bloch representation in a localized basis}
\label{sec:bloch}

\subsection{Lattice and reciprocal-space conventions}

Atomic units are used unless stated otherwise, with
$E_h\simeq4.359744722\times10^{-18}$~J and
$a_0\simeq5.291772105\times10^{-11}$~m.
One Rydberg is $E_h/2$, one electron volt is
$1.602176634\times10^{-19}$~J, one Angstrom is $10^{-10}$~m, and
one bar is $10^5$~Pa.

Let $\ba_1,\ba_2,\ba_3$ be the direct lattice vectors and
\begin{equation}
  \bR_{\bm{n}} = n_1\ba_1+n_2\ba_2+n_3\ba_3,\qquad
  \bm{n}\in\mathbb{Z}^3,
\end{equation}
an image-cell vector.  The reciprocal vectors satisfy
$\ba_i\cdot\bb_j=2\pi\delta_{ij}$.  \CPtwoK{} stores a k-point by the
fractional coordinate vector
\begin{equation}
  \bk = x_1\bb_1+x_2\bb_2+x_3\bb_3,
\end{equation}
so that the phase associated with image cell $\bm{n}$ is
\begin{equation}
  \exp(\ii\bk\cdot\bR_{\bm{n}})
  = \exp\!\left(2\pi\ii\,\bm{x}_{k}\cdot\bm{n}\right).
  \label{eq:phase}
\end{equation}
This fractional convention is used throughout the k-point source code and is
the default input convention for the \texttt{\&KPOINTS} section.

\subsection{Bloch sums and matrix elements}

Let $\phi_\mu(\br-\bR_{\bm{n}})$ denote a localized Gaussian basis function
translated to image cell $\bm{n}$.  The corresponding Bloch-adapted basis
function is
\begin{equation}
  \chi_{\mu\bk}(\br) =
  \sum_{\bm{n}} \exp\!\left(\ii\bk\cdot\bR_{\bm{n}}\right)
  \phi_\mu(\br-\bR_{\bm{n}}),
  \label{eq:blochsum}
\end{equation}
using cell-normalized Bloch sums. Dividing this expression by
$\sqrt{N_{\mathrm{cell}}}$ instead gives Born--von Karman supercell-normalized
functions. Per-cell densities would then carry an additional
$N_{\mathrm{cell}}$ factor. For any real-space matrix $A_{\mu\nu}^{\bm{n}}$ coupling basis
function $\mu$ in the home cell to basis function $\nu$ in image cell
$\bm{n}$, \CPtwoK{} forms the k-dependent matrix
\begin{equation}
  A_{\mu\nu}(\bk) =
  \sum_{\bm{n}} \exp\!\left(2\pi\ii\,\bm{x}_{k}\cdot\bm{n}\right)
  A_{\mu\nu}^{\bm{n}}.
  \label{eq:rs-to-k}
\end{equation}
Equation~\eqref{eq:rs-to-k} is used for the overlap matrix
$S(\bk)$, the KS matrix $H^\sigma(\bk)$, auxiliary-basis overlap matrices, and
the real-space objects needed by post-KS methods.  For the real, self-adjoint scalar operators considered here,
\begin{equation}
  A_{\mu\nu}^{\bm{n}} = A_{\nu\mu}^{-\bm{n}},
\end{equation}
and the k-space matrix is Hermitian,
\begin{equation}
  A_{\mu\nu}(\bk) = A_{\nu\mu}^{*}(\bk).
\end{equation}
The implementation therefore stores complex matrices as a pair comprising a
symmetric real part and an antisymmetric imaginary part whenever this symmetry
can be exploited.

\subsection{Consequences of localized nonorthogonal basis functions}

The Bloch transformation acts on sparse image-cell operators, not on an
operator initially represented in a dense reciprocal-space basis.  The object
transformed by Eq.~\eqref{eq:rs-to-k} is a finite set of sparse atom-pair blocks,
\begin{equation}
  \left\{A_{\mu\nu}^{\bm{n}}\right\}_{\bm{n}\in\mathcal{N}_{\mu\nu}},
  \qquad
  A_{\mu\nu}^{\bm{n}}=0
  \quad \mathrm{for}\quad
  \bm{n}\notin\mathcal{N}_{\mu\nu},
  \label{eq:finite-image-list}
\end{equation}
where the image-cell set $\mathcal{N}_{\mu\nu}$ is determined by Gaussian
extent, screening thresholds, cell shape, and the neighbor lists used by the
real-space GPW/GAPW machinery.  K-point sampling increases the number of transformed problems while
retaining this screened real-space representation. The retained image set
must be converged independently of the k-point quadrature.

The nonorthogonality of the Gaussian basis is equally important.  Every
k-point calculation involves the transformed overlap matrix $S(\bk)$ together
with the transformed KS matrix $H(\bk)$, so the eigenproblem is generalized
rather than ordinary.  Density matrices must then be transformed back to the
same image-cell representation that the Gamma-point code expects.  Schematically,
the grid density used by the GPW part is evaluated from the real-space matrix as
\begin{equation}
  \rho^\sigma(\br_g)
  =
  \sum_{\mu\nu}
  \sum_{\bm{m},\bm{n}}
  P_{\mu\nu}^{\sigma,\bm{n}}
  \phi_\mu(\br_g-\bR_{\bm{m}})
  \phi_\nu(\br_g-\bR_{\bm{m}}-\bR_{\bm{n}}),
  \label{eq:gpw-density-schematic}
\end{equation}
where $\br_g$ denotes a grid point and the sum over $\bm m$ implements
periodic accumulation. The inverse transform supplies the image-cell density required by
Eq.~\eqref{eq:gpw-density-schematic}. The existing GPW/GAPW density,
potential, and derivative routines can therefore consume the k-point
result through their real-space interfaces.

At every k-point and spin channel, the KS equations are solved as the
generalized Hermitian eigenproblem
\begin{equation}
  H^\sigma(\bk) C^\sigma(\bk)
  =
  S(\bk) C^\sigma(\bk) \varepsilon^\sigma(\bk),
  \label{eq:generalized-eigenproblem}
\end{equation}
with the metric normalization
\begin{equation}
  C^{\sigma\dagger}(\bk) S(\bk) C^\sigma(\bk) = I.
\end{equation}
For k-points whose fractional coordinates are restricted to $0$ or $1/2$ in
all periodic directions, the Bloch phases are real and a real-wavefunction path
can be used.  Otherwise \CPtwoK{} uses the complex path, represented internally
by separate real and imaginary full matrices.

\subsection{Density matrices and inverse transform}

The k-resolved density matrix is
\begin{equation}
  P_{\mu\nu}^{\sigma}(\bk)
  =
  \sum_{n} f_{n\bk}^{\sigma}
  C_{\mu n}^{\sigma}(\bk)
  C_{\nu n}^{\sigma *}(\bk),
  \label{eq:p-k}
\end{equation}
where $f_{n\bk}^{\sigma}$ is the occupation number. The electron count in
a nonorthogonal basis is
\begin{equation}
 N_{\mathrm{el}}=\sum_{\sigma\bk}w_{\bk}
 \Tr[P^\sigma(\bk)S(\bk)]
 =\sum_{\sigma n\bk}w_{\bk}f_{n\bk}^{\sigma}.
 \label{eq:metric-electron-count}
\end{equation}
This metric trace is an invariant check on both the Bloch transformation
and the symmetry reconstruction. With
$C=C^{\mathrm{R}}+\ii C^{\mathrm{I}}$, the real and imaginary parts stored by
the code are
\begin{align}
  P^{\mathrm{R}} &=
  C^{\mathrm{R}} f C^{\mathrm{R}T}
  + C^{\mathrm{I}} f C^{\mathrm{I}T},
  \label{eq:p-real}\\
  P^{\mathrm{I}} &=
  C^{\mathrm{I}} f C^{\mathrm{R}T}
  - C^{\mathrm{R}} f C^{\mathrm{I}T}.
  \label{eq:p-imag}
\end{align}
The real-space density matrix used by the GPW/GAPW machinery is then obtained
by the weighted inverse transform
\begin{equation}
  P_{\mu\nu}^{\sigma,\bm{n}}
  =
  \sum_{\bk} w_{\bk}
  \exp\!\left(-2\pi\ii\,\bm{x}_{k}\cdot\bm{n}\right)
  P_{\mu\nu}^{\sigma}(\bk).
  \label{eq:k-to-rs}
\end{equation}
For scalar collinear Hamiltonians with complex-conjugation symmetry and a
symmetry-closed mesh, the imaginary contributions cancel in the full sum.
The real image-cell density is therefore accumulated as
\begin{equation}
  P_{\mu\nu}^{\sigma,\bm{n}}
  =
  \sum_{\bk} w_{\bk}
  \left[
  P_{\mu\nu}^{\sigma,\mathrm{R}}(\bk)
  \cos(2\pi\bm{x}_{k}\cdot\bm{n})
  +
  P_{\mu\nu}^{\sigma,\mathrm{I}}(\bk)
  \sin(2\pi\bm{x}_{k}\cdot\bm{n})
  \right],
  \label{eq:k-to-rs-real}
\end{equation}
with the appropriate sign changes for transposed blocks stored through sparse
matrix symmetry.  The same transformation is used for the energy-weighted
density matrix
\begin{equation}
  W_{\mu\nu}^{\sigma}(\bk)
  =
  \sum_n f_{n\bk}^{\sigma}\varepsilon_{n\bk}^{\sigma}
  C_{\mu n}^{\sigma}(\bk)
  C_{\nu n}^{\sigma *}(\bk),
  \label{eq:w-k}
\end{equation}
which enters Pulay-force terms when the overlap metric is not the unit matrix.

\section{Symmetry reduction and real wavefunctions}
\label{sec:k290-spglib}

Symmetry reduction connects a compact
Brillouin-zone quadrature and a usable localized-basis implementation.  It
reduces the number of self-consistent k-point eigenproblems, but it must also
preserve the operations required to expand k-dependent matrices back over the
full orbit.  This distinction is especially important in \Quickstep, where the
primary stored objects are image-cell matrices rather than plane-wave
coefficients labelled directly by $\bk+\bG$.  The symmetry operation must
therefore act simultaneously on the crystal, the atom ordering, the local
angular-momentum shells, the image-cell vector, and the Bloch phase.
Only symmetries of the electronic Hamiltonian are admissible. Atomic kinds,
magnetic order, external fields, and occupations can reduce this group
relative to the crystallographic space group.

For generated grids, crystal symmetries are obtained from the same symmetry
machinery used elsewhere in \CPtwoK.  If \texttt{SYMMETRY} is active, the
atomic kinds and scaled coordinates are used to find the space-group operations
within the tolerance \texttt{EPS\_SYMMETRY}. With \texttt{SYMMETRY F},
regular meshes still use time-reversal pairing by default. A strict
unreduced reference requires both \texttt{SYMMETRY F} and
\texttt{FULL\_GRID T}. The symmetry data associated
with each k-point include the number of equivalent points, the reduced and
full-grid coordinates, rotation matrices, atom permutations, and phase factors.
The two input choices for symmetry search and mesh reduction are
\begin{align}
  \texttt{SYMMETRY\_BACKEND} &\in \{\texttt{K290},\texttt{SPGLIB}\},\\
  \texttt{SYMMETRY\_REDUCTION\_METHOD} &\in
  \{\texttt{K290},\texttt{SPGLIB}\}.
\end{align}
The default is the K290 backend. Its crystal-symmetry routines, implemented in
\texttt{kpsym.F}, derive from the group-theoretical programs of Worlton and
Warren~\cite{Worlton1972,Warren1974}. They determine the crystallographic and
atomic symmetry operations, while reduction of the generated k-point mesh is
performed separately in \texttt{cryssym.F}. The resulting K290 operations and
atom mappings are stored for subsequent symmetry transformations in
\CPtwoK{}. The SPGLIB backend, enabled at build time through
\texttt{CP2K\_USE\_SPGLIB} and the preprocessor symbol \texttt{\_\_SPGLIB},
obtains space-group operations from SPGLIB, including fractional
translations~\cite{Togo2024}. These operations are exposed through the
\texttt{spglib\_f08.F} Fortran interface and are used for k-point reduction
and subsequent symmetry transformations.

The tolerance \texttt{EPS\_SYMMETRY} has a backend-specific geometric meaning.
K290 applies it in lattice and fractional-coordinate checks, whereas SPGLIB
uses a Cartesian distance tolerance in bohr.

A direct-space symmetry operation can be written as
\begin{equation}
  g = (R_g|\bm{\tau}_g),
  \qquad
  \bm s \mapsto R_g\bm s+\bm{\tau}_g ,
\end{equation}
where $\bm s$ is a fractional atomic coordinate, with integer
fractional-coordinate rotation $R_g$ and fractional translation
$\bm{\tau}_g$.  The corresponding reciprocal-space action is
\begin{equation}
  \bm x_k \mapsto R_g^{-T}\bm x_k + \bm m_g ,
  \label{eq:k-sym-action}
\end{equation}
where $\bm m_g\in\mathbb Z^3$ folds the result back into
the selected k-point mesh.  K290 and SPGLIB differ primarily in how the allowed
operations and k-point orbits are found. Both must provide an operation
that connects every reducible
k-point to its representative and can be used consistently for atom
permutations, basis rotations\cite{Choi1999}, and phase factors in the matrix transformations.
For a mesh $\mathcal{K}$ invariant under the admitted symmetry group
$\mathcal{G}$, the orbit of
a k-point is
\begin{equation}
  \mathcal{O}(\bk) =
  \left\{
  B(R_g^{-T}\bm x_k+\bm m_g) \;|\; g\in\mathcal{G}
  \right\}
  \cap \mathcal{K}.
  \label{eq:k-orbit}
\end{equation}
Here $B=(\bb_1,\bb_2,\bb_3)$. If time reversal is admitted, its additional
minus sign is included in the reciprocal action. An irreducible mesh chooses
one representative $\bar{\bk}$ from each orbit.
For an equally weighted full mesh, the representative weight is
\begin{equation}
  w_{\bar{\bk}} =
  \frac{|\mathcal{O}(\bar{\bk})|}{|\mathcal{K}|},
  \qquad
  \sum_{\bar{\bk}\in\mathcal{K}_{\mathrm{irr}}}w_{\bar{\bk}}=1.
  \label{eq:irr-weight}
\end{equation}
This is the simple reason why a symmetry-reduced calculation can reproduce a
full-mesh Brillouin-zone integral. The missing k-points are not discarded, but
are represented by the orbit multiplicity in Eq.~\eqref{eq:irr-weight}.

For scalar band quantities this orbit logic is sufficient,
\begin{equation}
  \sum_{\bk\in\mathcal{K}} \frac{1}{|\mathcal{K}|} F(\bk)
  =
  \sum_{\bar{\bk}\in\mathcal{K}_{\mathrm{irr}}}
  w_{\bar{\bk}}F(\bar{\bk}),
  \label{eq:scalar-orbit-average}
\end{equation}
provided that $F(g\bk)=F(\bk)$.  Matrix quantities require more information
because the basis functions themselves are rotated, permuted, and phase
shifted by $g$. To make the phase explicit, let atom $I$ map to atom
$p_g(I)$ with an integer image shift $\bm n_{gI}$ such that
\begin{equation}
 R_g\bm s_I+\bm\tau_g=\bm s_{p_g(I)}+\bm n_{gI}.
 \label{eq:atom-symmetry-image}
\end{equation}
For a unitary operation and the lattice-periodic basis of
Eq.~\eqref{eq:blochsum}, the coefficient map has blocks
\begin{equation}
 [T_g(\bk)]_{p_g(I)b,Ia}
 =e^{-2\pi\ii\,\bm x_{gk}\cdot\bm n_{gI}}\,
 D^{(I)}_{ba}(g),
 \label{eq:ao-symmetry-block}
\end{equation}
where $a,b$ label local orbitals and $D^{(I)}(g)$ rotates their
angular components. The translation in a nonsymmorphic operation enters
through Eq.~\eqref{eq:atom-symmetry-image}. Dropping the image shift
therefore changes the Bloch representation. An integer reciprocal folding
does not add an atomic-position phase in this basis convention.

Let $T_g$ map coefficient vectors at $\bk$ to those at $g\bk$,
and let $\mathcal K_g$ denote complex conjugation for an antiunitary
operation and the identity otherwise. The density matrix transforms as
\begin{equation}
  P(g\bk)=T_g(\bk)\,\mathcal K_g[P(\bk)]\,T_g^\dagger(\bk),
  \label{eq:symmetry-matrix-transform}
\end{equation}
where $T_g$ contains the atom permutation, angular-momentum rotation, and
lattice phase. Nonorthogonality requires the corresponding metric relation
\begin{equation}
 T_g^\dagger S(g\bk)T_g=\mathcal K_g[S(\bk)],\qquad
 T_g^\dagger H(g\bk)T_g=\mathcal K_g[H(\bk)].
 \label{eq:metric-covariance}
\end{equation}
Thus density and operator matrices must not be assigned the same generic
similarity transform without specifying the coefficient convention.
The metric relation also guarantees
\begin{equation}
 \Tr[P(g\bk)S(g\bk)]
 =\mathcal K_g\{\Tr[P(\bk)S(\bk)]\}
 =\Tr[P(\bk)S(\bk)],
 \label{eq:symmetry-charge-invariant}
\end{equation}
because the physical electron count is real. The density returned to real space
therefore involves the symmetry-expanded sum
\begin{equation}
  P^{\bm{n}} =
  \sum_{\bar{\bk}\in\mathcal{K}_{\mathrm{irr}}}
  \frac{1}{|\mathcal{K}|}
  \sum_{\bk'\in\mathcal{O}(\bar{\bk})}
  \exp(-\ii\bk'\cdot\bR_{\bm{n}})
  P(\bk').
  \label{eq:symmetry-expanded-density}
\end{equation}
Equations~\eqref{eq:symmetry-matrix-transform} and
\eqref{eq:symmetry-expanded-density} are the software-critical part of
symmetry reduction. The code must know not only how many k-points are
equivalent, but also how the corresponding matrix representation transforms.

The two selectors separate this reduction/backend logic.  If
\texttt{SYMMETRY\_BACKEND} is set explicitly and
\texttt{SYMMETRY\_REDUCTION\_METHOD} is omitted, the reduction method follows
the backend.  \texttt{SYMMETRY\_BACKEND SPGLIB} requires
\texttt{SYMMETRY\_REDUCTION\_METHOD SPGLIB}, because the SPGLIB backend uses
SPGLIB operations for both generating the irreducible set and applying the
transformations. A mixed diagnostic mode combines
\texttt{SYMMETRY\_REDUCTION\_METHOD SPGLIB} with
\texttt{SYMMETRY\_BACKEND K290}.  In that case SPGLIB proposes the k-point
orbits, but only mappings represented by the K290 backend are accepted for the
actual transformations.  This is useful for comparing reductions and for
guarding against symmetry mappings that cannot be expressed in the K290
operation set.

The same logic applies to regular Monkhorst--Pack and MacDonald meshes and, in
the current code, also to \texttt{SCHEME GENERAL} when the explicit list is
equally weighted and closed under the selected symmetry operations.  For band
paths or intentionally nonuniform explicit k-point lists, symmetry should
remain disabled because the order and weights are part of the requested
property.  \texttt{INVERSION\_SYMMETRY\_ONLY} restricts the reduction to the
time-reversal/inversion pairing
\begin{equation}
  \varepsilon_{n\bk} = \varepsilon_{n,-\bk},
\end{equation}
provided the scalar, collinear Hamiltonian admits complex-conjugation
symmetry. This pairing does not require spatial inversion of the crystal
and must not be generalized to arbitrary magnetic or spinor Hamiltonians.
For regular meshes, \texttt{SYMMETRY T FULL\_GRID T} suppresses atomic
reduction but still retains this pairing. Only the combination
\texttt{SYMMETRY F FULL\_GRID T} is a strict full-mesh reference.

In the version used for the numerical tests, atomic-symmetry reduction
requires orthogonal cells in CP2K's canonical cell convention. Other cells
use the guarded time-reversal path. These implementation conditions do not
restrict the general covariance relations above. Atomic configurations and finite integration grids
can also lower the effective symmetry. Consequently, backend agreement in
orbit counts alone is insufficient. Energies, forces, stress, and the
metric covariance in Eq.~\eqref{eq:metric-covariance} must be checked against
the equivalent full mesh. Geometry changes require updated symmetry data,
so an initial crystal space group is not a permanent constraint on a trajectory.

The real-wavefunction option is an optimization rather than a different
physical approximation.  It is valid when all Bloch phase factors are real,
which requires fractional components compatible with $0$ or $1/2$ in the
periodic directions.  If this condition is not met, the complex path is the
proper representation.  The default in the input is complex, reflecting the
safe general case.

\section{Real--k transformations}
\label{sec:transforms}

\subsection{Sparse production transform}

The production transformation in \texttt{kpoint\_methods.F} iterates over the
neighbor-list image cells that contain nonzero atom-block pairs.  For a cell
$\bm{n}$, it evaluates
\begin{equation}
  c_{\bk\bm{n}}=\cos(2\pi\bm{x}_{k}\cdot\bm{n}),\qquad
  s_{\bk\bm{n}}=\sin(2\pi\bm{x}_{k}\cdot\bm{n})
\end{equation}
and accumulates
\begin{align}
  A^{\mathrm{R}}(\bk) &\leftarrow
  A^{\mathrm{R}}(\bk)+c_{\bk\bm{n}}A^{\bm{n}},\\
  A^{\mathrm{I}}(\bk) &\leftarrow
  A^{\mathrm{I}}(\bk)+s_{\bk\bm{n}}A^{\bm{n}}.
\end{align}
For real wavefunctions the imaginary accumulation is skipped.  Symmetric sparse
storage requires special handling because a stored block may represent either
$A_{\mu\nu}^{\bm{n}}$ or its transposed partner
$A_{\nu\mu}^{-\bm{n}}$. The code applies the corresponding sign convention so
that the final k-space matrix is Hermitian.

The inverse transform uses the k-point weights and the negative phase in
Eq.~\eqref{eq:k-to-rs}.  If a reduced symmetry grid is used, density matrices
are expanded over the symmetry-equivalent k-points before accumulation.  The
current K290/SPGLIB layer provides the representative, operation, atom
permutation, and phase information needed for this expansion.  Whenever a
downstream interface cannot apply the selected symmetry mapping consistently,
the code aborts with a diagnostic or falls back to a full-grid path rather than
silently mixing incompatible matrix representations.  For nonuniform paths and
for validation of new post-processing interfaces, \texttt{FULL\_GRID} therefore
remains the conservative setting.

\subsection{Array transform}

For calculations where the real--k transform is not performance critical, the
array-based transformation module provides dense-array variants.  These
routines first replicate sparse image-cell matrices into four-dimensional
Fortran arrays indexed by spin, row, column, and image cell.  The forward
transform is the direct sum
\begin{equation}
  A(\bk) = \sum_{i_{\mathrm{image}}}
  \exp\!\left(2\pi\ii\,\bm{x}_{k}\cdot\bm{n}_{i}\right)
  A^{i_{\mathrm{image}}},
\end{equation}
with an optional derivative factor
\begin{equation}
  \frac{\partial A(\bk)}{\partial k_{\alpha}}
  =
  \sum_{\bm{n}}
  \ii R_{\bm{n},\alpha}
  \exp(\ii\bk\cdot\bR_{\bm{n}})
  A^{\bm{n}}.
\end{equation}
The inverse helper adds a single k-point matrix to all image cells through
Eq.~\eqref{eq:k-to-rs-real}.  These routines are easier to reuse in
post-processing and response code, but they trade memory for simplicity.

\subsection{Fast lattice transforms}

Complete uniformly shifted meshes permit a fast Fourier transform (FFT).
Write $\bm x_{\bm m}=\bm m/\bm N+\bm\delta$, with componentwise division.
The image-cell coefficients are first folded into a mesh-sized array,
\begin{equation}
 B_{\bm\ell}=\sum_{\bm n\equiv\bm\ell\;\mathrm{mod}\;\bm N}
 e^{2\pi\ii\bm\delta\cdot\bm n}A^{\bm n},\qquad
 A(\bm x_{\bm m})=\sum_{\bm\ell}B_{\bm\ell}
 e^{2\pi\ii\sum_\alpha m_\alpha\ell_\alpha/N_\alpha}.
 \label{eq:lattice-fft}
\end{equation}
The folding is exact for the sampled points, not a truncation of the physical
image list. The shift phase must be applied before folding, since different
images with the same remainder can carry different phases. Derivatives
similarly require multiplication by the original image vector before folding.
For each matrix batch, the transform cost changes from a direct
$O(N_kN_{\mathrm{image}})$ phase sum to folding plus
$O(N_k\log N_k)$ operations, at the expense of mesh buffers.
This concerns the transform, not the total electronic-structure scaling.

\texttt{LATTICE\_FFT AUTO}, the default, selects eligible complete regular
meshes with at least 27 points subject to layout and memory checks.
\texttt{ON} bypasses this size heuristic, not the eligibility guards.
\texttt{OFF} selects the direct reference path. Irregular or incomplete
lists retain direct transforms. The mathematical full-grid FFT must not be
confused with a Fourier transform of irreducible representatives alone,
which omits the matrix symmetry expansion in
Eq.~\eqref{eq:symmetry-expanded-density}.

\section{Self-consistency and orbital optimization}
\label{sec:scf}

In each diagonalization-based SCF step, \CPtwoK{}
constructs real-space KS and overlap matrices as in a Gamma calculation.  For
each k-point needed by a local k-point group, these matrices are transformed
according to Eq.~\eqref{eq:rs-to-k}.  The transformed real and imaginary parts
are copied from sparse block matrices into the block-cyclic dense matrices used
by the parallel eigensolvers.  The code then solves
Eq.~\eqref{eq:generalized-eigenproblem}, stores real and imaginary molecular
orbital coefficients separately, determines occupations from the complete
k-resolved eigenvalue spectrum, builds the k-space density matrices using
Eqs.~\eqref{eq:p-real}--\eqref{eq:p-imag}, and finally transforms them back to
real-space image-cell matrices using Eq.~\eqref{eq:k-to-rs-real}.  These
real-space density matrices are then consumed by the usual GPW/GAPW density,
potential, and energy routines.

The occupations obey the global electron count in
Eq.~\eqref{eq:metric-electron-count}.
The implementation gathers the eigenvalues across k-point groups before
occupation assignment.  Metallic occupations can be obtained with
Fermi--Dirac smearing, Gaussian smearing, first-order
Methfessel--Paxton smearing, and Marzari--Vanderbilt cold
smearing~\cite{MethfesselPaxton1989,Marzari1999,dosSantos2023}.  For
Fermi--Dirac smearing the occupation has the form
\begin{equation}
  f_{n\bk}^{\sigma}
  =
  \frac{f_{\max}^{\sigma}}
  {1+\exp[(\varepsilon_{n\bk}^{\sigma}-\mu)/k_{\mathrm{B}}T_{\mathrm{el}}]},
  \label{eq:fd}
\end{equation}
where the chemical potential $\mu$ is chosen so that
Eq.~\eqref{eq:metric-electron-count} is satisfied.  The code also supports fixed
magnetization in the smearing machinery by applying the occupation constraint
separately to spin channels.

Explicit Bloch eigenspaces are not the only representation of the electronic
density. Fermi-operator decompositions evaluate the finite-temperature
density matrix directly~\cite{Ceriotti2008}. Their extension to self-consistent
forces and molecular dynamics was developed by Richters and
K\"uhne~\cite{Richters2014}. Nonorthogonal local submatrix methods provide a
complementary locality-based route to large-scale calculations~\cite{Schade2022}.
These methods concern the construction of the density matrix, whereas the
present symmetry reduction concerns its reciprocal-space representation and
quadrature. A density-matrix approximation alone does not supply the
gauge-consistent full-mesh orbital frames and neighboring overlaps required
for the Wannier analysis below.

\subsection{Complex orbital optimization and iterative solvers}

The current implementation is not restricted to full diagonalization.
Complex orbital transformation (OT) extends the constrained minimization
of Ref.~\cite{VandeVondele2003} to weighted k-point sets. For a reference
coefficient matrix $C_0$ and a tangent displacement $X$, an orthonormal
parameterization is
\begin{equation}
 C(X)=C_0\cos Q+X\,\operatorname{sinc}Q,\qquad
 Q=(X^\dagger SX)^{1/2},\qquad C_0^\dagger SX=0,
 \label{eq:complex-ot}
\end{equation}
where $\operatorname{sinc}Q=\sin(Q)/Q$ is defined by continuity at zero.
Together with $C_0^\dagger SC_0=I$, this gives $C(X)^\dagger SC(X)=I$.
All matrices and constraints in this equation refer to the same k-point.
Complex conjugate transposes, rather than real transposes, are essential.
The implementation provides conjugate-gradient, direct-inversion, and
limited-memory optimization variants with explicit overlap-dependent
preconditioning. No single preconditioner is uniformly optimal.

For Fermi--Dirac occupations, finite-temperature OT minimizes the
fixed-electron-number Mermin free energy~\cite{Mermin1965},
\begin{align}
 \mathcal F[C,f]&=E[C,f]-T_{\mathrm{el}}\mathcal S[f],\\
 \mathcal S[f]&=-k_{\mathrm B}\sum_{\sigma n\bk}w_{\bk}f_{\max}^{\sigma}
 \left[p_{n\bk}^{\sigma}\ln p_{n\bk}^{\sigma}
 +(1-p_{n\bk}^{\sigma})\ln(1-p_{n\bk}^{\sigma})\right],\qquad
 p=f/f_{\max}.
 \label{eq:mermin-entropy}
\end{align}
The spin degeneracy is absorbed in $f_{\max}$. Separate collinear spin
channels have $f_{\max}=1$. Rotations within the optimized subspace become
variational when occupations differ. Auxiliary orbital energies control
the occupations, and a common chemical potential enforces
Eq.~\eqref{eq:metric-electron-count}. Thus \texttt{ROTATION} and \texttt{ENERGIES}
are both needed for direct OT with smearing. In the linearized occupation
response, with $a_i=\partial f_i/\partial\varepsilon_i$ and $i$ denoting
band, spin, and k-point, fixed electron number requires
\begin{equation}
 \delta\mu=\frac{\sum_iw_i a_i\,\delta\varepsilon_i}{\sum_iw_i a_i},
 \qquad \delta f_i=a_i(\delta\varepsilon_i-\delta\mu).
 \label{eq:fixed-n-response}
\end{equation}
This expression applies when the denominator is nonzero. The zero-temperature
gapped limit is treated as a fixed-occupation problem. It explains why
occupation response and its preconditioner must communicate across k-point
groups rather than independently optimize electron numbers at each point.
\texttt{ADDED\_MOS AUTO} grows the virtual-state buffer when its upper edge
is appreciably occupied, but cannot compensate for an insufficient AO basis.
Forces and stress at finite temperature must be compared with derivatives of
$\mathcal F$, not of the uncorrected electronic energy. Other smearing
distributions use their corresponding generalized free-energy corrections.

In contrast, \texttt{SCF/DIAGONALIZATION/OT} solves a fixed-Hamiltonian
eigenspace problem inside the conventional density-mixing loop. It does not
optimize the additional direct-OT occupation variables. Complex blocked
Davidson likewise iterates on the residual
\begin{equation}
 R=HC-SC\varepsilon,\qquad
 X_{\mathrm{new}}=\mathcal P R,
\end{equation}
followed by metric orthogonalization and a projected generalized
eigenproblem. The shared operator service assembles and applies the same
$H(\bk)$ and $S(\bk)$ for these consumers, reducing the risk of solver-specific
phase or distribution inconsistencies. Solver availability does not imply
universal support for every functional or response calculation. The Supporting
Information records method-specific compatibility conditions.

\subsection{Orbital-dependent DFT+U on k-point meshes}
\label{sec:dft-plus-u}

Localized orbitals also permit a Hubbard correction during k-point SCF.
For the L\"owdin variant, let $P^{\sigma}(\bk)$ denote a density
matrix normalized to one electron per occupied spin orbital, and let
$\mathcal C_I$ contain the selected AOs on atom $I$. Their occupations are
\begin{equation}
 n_{I\mu}^{\sigma}=\sum_{\bk}w_{\bk}
 [S^{1/2}(\bk)P^{\sigma}(\bk)S^{1/2}(\bk)]_{\mu\mu},
 \qquad \mu\in\mathcal C_I.
 \label{eq:hubbard-lowdin-occupation}
\end{equation}
A restricted calculation has two identical spin channels. For the
orbital-diagonal Dudarev-type term, with the separate Hund coupling
suppressed for clarity, the energy and orbital potential are~\cite{Dudarev1998}
\begin{equation}
 E_U=\frac12\sum_{I\sigma}U_I^{\mathrm{eff}}
 \sum_{\mu\in\mathcal C_I}
 (n_{I\mu}^{\sigma}-(n_{I\mu}^{\sigma})^2),
 \qquad
 V_{I\mu}^{\sigma}=U_I^{\mathrm{eff}}
 (\tfrac12-n_{I\mu}^{\sigma}).
 \label{eq:hubbard-lowdin-energy}
\end{equation}
CP2K retains the selected columns $B(\bk)$ of $S^{1/2}(\bk)$
and applies $\Delta H^{\sigma}(\bk)=B(\bk)\,\mathrm{diag}
(V^{\sigma})B^{\dagger}(\bk)$ directly to the Bloch Hamiltonian.
The k-weighted occupations are reduced across k-point groups and mixed
consistently with the SCF density. Analytic forces and stress additionally
require the variation of the overlap square root, determined by
$S^{1/2}\,\delta S^{1/2}+\delta S^{1/2}\,S^{1/2}=\delta S$.
Its contribution enters the energy-weighted density matrix before the
usual real-space overlap derivatives are evaluated.

The L\"owdin path supports standard diagonalization and direct OT on
full meshes or with time-reversal-only reduction. Atomic point-group
reduction and enforced occupations are guarded. It is therefore a
connected k-point capability, not a validation of the atomic-symmetry
reduction tested below.

\section{Implementation and data flow}
\label{sec:implementation}

\subsection{Operators and data structures}

The implementation separates symmetry metadata, image-cell operators,
and reciprocal-space solvers. The Supporting Information maps these roles
to source modules.
The main type, \texttt{kpoint\_type}, stores input choices, the generated
k-point list and weights, symmetry metadata, image-cell maps, and the
communicators that define the k-point parallel layout.  For each locally owned
k-point, \texttt{kpoint\_env\_type} stores molecular orbitals (MOs), density
matrices, energy-weighted matrices, overlap-related matrices, and auxiliary
basis objects.  Complex data are represented by an explicit first dimension
with index 1 for the real part and index 2 for the imaginary part.  This design
keeps the interface close to the existing real-valued matrix infrastructure
while allowing the dense eigensolvers to operate on complex matrices when
needed.

The initialization sequence is deliberately layered.  First, input values are
read and a k-point grid is generated.  Second, the global communicator is split
into k-point groups.  Third, molecular orbital containers and matrix pools are
allocated in the local k-point groups.  Fourth, neighbor-list image cells are
compressed into integer maps
\begin{equation}
  \bm{n}\longleftrightarrow i_{\mathrm{image}},
\end{equation}
so that sparse image-cell matrices can be transformed with minimal lookup
overhead.  The routine ordering in a standard \Quickstep{} initialization is
therefore
\begin{equation}
\begin{split}
  &\texttt{read\_kpoint\_section}
  \rightarrow
  \texttt{kpoint\_initialize}
  \rightarrow
  \texttt{kpoint\_env\_initialize} \\
  &\rightarrow
  \texttt{kpoint\_initialize\_mos}
  \rightarrow
  \texttt{wfi\_create\_for\_kp},
\end{split}
\end{equation}
followed by the usual construction of real-space matrices and density objects.

\subsection{K-point sets and input conventions}

The input section \texttt{\&KPOINTS} below
\texttt{\&FORCE\_EVAL/\&DFT} supports the five schemes
\texttt{NONE}, \texttt{GAMMA}, \texttt{MONKHORST-PACK},
\texttt{MACDONALD}, and \texttt{GENERAL}.  With no explicit section,
\texttt{NONE} is used and the standard Gamma-only code path is retained.
\texttt{GAMMA} creates a one-point grid at $\bm{x}_{k}=0$ with unit weight.
\texttt{MONKHORST-PACK} takes three integers and delegates grid
construction~\cite{Monkhorst1976}, optional symmetry reduction, and weight
generation to the crystal-symmetry machinery.  \texttt{MACDONALD} uses the
same grid dimensions together with an explicit shift vector following the
generalization by MacDonald~\cite{MacDonald1978}.  \texttt{GENERAL} reads
repeated lines
\begin{verbatim}
KPOINT  kx  ky  kz  weight
\end{verbatim}
and normalizes the supplied weights so that $\sum_{\bk}w_{\bk}=1$.

For generated paths such as band structures, a separate repeated
\texttt{\&KPOINT\_SET} section stores named special points and a number of
interpolating points.  If $M$ special points are connected by $N_{\mathrm{seg}}$
points per segment, the number of generated k-points is
\begin{equation}
  N_k = (M-1)N_{\mathrm{seg}}+1.
\end{equation}
The interpolation is linear in fractional reciprocal coordinates.  For
Monkhorst--Pack meshes constructed explicitly in the band-structure utilities,
the fractional coordinate in direction $\alpha$ is
\begin{equation}
  x_{\alpha}(m) =
  \frac{2m-N_{\alpha}-1}{2N_{\alpha}},
  \qquad m=1,\ldots,N_{\alpha}.
  \label{eq:mp-grid}
\end{equation}

The keyword \texttt{UNITS} controls the interpretation of explicitly supplied
k-point coordinates.  In \texttt{B\_VECTOR} mode the numbers are the
fractional components $x_\alpha$.  In Cartesian modes the input vector
$\bk_{\mathrm{cart}}$ is converted to fractional coordinates through
\begin{equation}
  x_\alpha =
  \frac{\bk_{\mathrm{cart}}\cdot\ba_\alpha}{2\pi},
  \label{eq:cart-to-frac}
\end{equation}
for a physical Cartesian wavevector. The numerical input in
\texttt{CART\_BOHR} and \texttt{CART\_ANGSTROM} is expressed in
$2\pi/\mathrm{bohr}$ and $2\pi/\mathrm{\mathring A}$, respectively.
It must be multiplied by $2\pi$ before applying this physical-wavevector
formula. The flags
\texttt{SYMMETRY} and \texttt{FULL\_GRID} determine whether crystal symmetries
are used to reduce the grid and whether the unreduced grid is retained.
\texttt{SYMMETRY\_BACKEND} and \texttt{SYMMETRY\_REDUCTION\_METHOD} select the
K290 or SPGLIB symmetry path discussed in Section~\ref{sec:k290-spglib}, while
\texttt{INVERSION\_SYMMETRY\_ONLY} restricts the reduction to
time-reversal/inversion symmetry.  The current tolerance keyword is
\texttt{EPS\_SYMMETRY}. The older \texttt{EPS\_GEO} spelling is still accepted
as an alias.
\texttt{WAVEFUNCTIONS REAL} requests real wavefunctions. The code checks that
the k-point coordinates are compatible with real Bloch phases and otherwise
warns the user.  \texttt{PARALLEL\_GROUP\_SIZE} controls the decomposition
over k-points and is discussed in Section~\ref{sec:parallel}.

\subsection{Parallel decomposition}
\label{sec:parallel}

K-points provide a natural outer level of parallelism in addition to the dense
linear algebra within each k-point. Let $P$ be the number of
Message Passing Interface (MPI) ranks,
$g$ the number of ranks assigned to one k-point group, and
$G=P/g$ the number of k-point groups. The implementation requires
\begin{equation}
  P \bmod g = 0,\qquad G \leq N_k,
  \label{eq:parallel-constraints}
\end{equation}
so that every group owns at least one contiguous range of k-points. The
k-points are distributed as evenly as possible among the groups, such that
\begin{equation}
  N_{k,i}^{\mathrm{local}}
  \in
  \left\{
    \left\lfloor \frac{N_k}{G} \right\rfloor,
    \left\lceil \frac{N_k}{G} \right\rceil
  \right\},
\end{equation}
where the numbers of k-points assigned to any two groups differ by at most
one.

The keyword \texttt{PARALLEL\_GROUP\_SIZE} selects $g$. A value of $-1$
chooses the smallest group size that satisfies
Eq.~\eqref{eq:parallel-constraints}, thereby maximizing the number of
k-point groups. A value of $0$ assigns all MPI ranks to a single k-point
group and disables k-point parallelism. A positive value requests the
corresponding group size, subject to the constraints in
Eq.~\eqref{eq:parallel-constraints}.

Internally \CPtwoK{} creates a two-dimensional Cartesian communicator. One
dimension spans the ranks within a k-point group and is used for distributed
dense eigensolvers and matrix copies. The other dimension connects equivalent
ranks across k-point groups and is used for communication over k-points, such as
the collection of eigenvalues for occupation assignment and the combination of
k-point contributions to density matrices. This layout keeps the dense linear
algebra within a k-point group while still allowing global k-weighted quantities
to be reduced or redistributed across groups without duplicating the underlying
communication logic in high-level algorithms.

The implementation also provides thread-level parallelism over k-points when
a k-point group contains a single MPI rank. In this case, the k-points assigned
to that group can be processed concurrently by OpenMP threads, since no
intra-group distributed matrix operations are required. Thread-private operator
and dense-linear-algebra workspaces are used to keep the per-k-point
calculations independent. This execution mode applies both to pure OpenMP
calculations and to hybrid MPI/OpenMP calculations with one MPI rank per
k-point group. Threading in the underlying dense linear-algebra libraries is
typically limited in this mode to avoid nested oversubscription.

\subsection{Restart and output}

K-point restarts are written as unformatted binary files with extension
\texttt{.kp}.  The file header stores a version number, atom and basis counts,
the number of basis sets and shells, and the number of spherical functions per
shell.  For each spin and image cell, the restart stores the image-cell vector
and the corresponding dense representation of the real-space density matrix.
On read, the file image-cell vectors are mapped back onto the current
\texttt{cell\_to\_index} table.  If a restart contains image cells not present
in the current neighbor-list representation, they are ignored. If the number
of atoms or basis functions is inconsistent, the read path either reports a
restart mismatch or aborts, depending on the inconsistency.

For post-processing, \texttt{kpoint\_mo\_dump.F} writes a text file containing
k-resolved eigenvalues, occupations, real and imaginary molecular-orbital
coefficients, and optionally overlap matrices.  The header includes cell
vectors, atom positions, atom-to-orbital mappings, k-point coordinates and
weights, and either a Gaussian basis definition or an angular-momentum map.
This is useful for projected density-of-states analysis, external band
post-processing, and regression comparisons against independent tools.

\section{Gauge-consistent orbital analysis}
\label{sec:analysis}

The reconstruction of a density matrix is not the final use of the
reciprocal-space representation. Wannier construction requires overlaps
between states at different k-points.
Orbital export and response calculations impose additional constraints
on atomic images, band ordering, and phase conventions. This section
develops these applications from the same localized-basis operators.

\subsection{Full-mesh data and orbital reconstruction for Wannier90}

The Wannier90 interface~\cite{Mostofi2008,Pizzi2020} is a particularly sensitive consumer of the k-point
layer because Wannier90 expects a complete mesh, a consistent nearest-neighbor
k-point topology, and overlap matrices between cell-periodic Bloch functions.
The CP2K print section \texttt{\&DFT/\&PRINT/\&WANNIER90} writes the
\texttt{seed.win}, \texttt{seed.mmn}, and \texttt{seed.eig} files and, for a
complete-band Bloch-phase export, an explicit identity-projection
\texttt{seed.amn} file.  The controlling keywords include
\texttt{SEED\_NAME}, \texttt{MP\_GRID}, \texttt{KPOINTS\_SOURCE},
\texttt{REUSE\_SCF\_MOS}, \texttt{VALIDATE\_REUSE\_SCF\_MOS},
\texttt{USE\_BLOCH\_PHASES}, \texttt{ADDED\_MOS}, \texttt{EXCLUDE\_BANDS},
and \texttt{WANNIER\_FUNCTIONS}.

Two regular-mesh k-point sources are implemented. With \texttt{KPOINTS\_SOURCE MP\_GRID},
CP2K retains the historical behavior by constructing a full Monkhorst--Pack
mesh from \texttt{WANNIER90\%MP\_GRID}, disabling symmetry for the export, and
diagonalizing the full mesh used by Wannier90. With
\texttt{KPOINTS\_SOURCE SCF}, the export mesh is derived from the active
\texttt{\&KPOINTS} section of the SCF calculation.  Gamma, Monkhorst--Pack,
MacDonald, and explicit \texttt{GENERAL} meshes are supported.  If the SCF mesh
was symmetry-reduced by K290 or SPGLIB, the Wannier90 export regenerates the
corresponding unreduced mesh because the neighbor graph returned by
Wannier90's setup step is defined on the full grid.  Explicit
\texttt{GENERAL} meshes must be equally weighted. If \texttt{MP\_GRID} is not
given, CP2K attempts to infer the grid dimensions and aborts if the list cannot
be interpreted as a full mesh.

After Wannier90 has generated the neighbor list, CP2K writes the overlap
matrices
\begin{equation}
  M_{mn}^{(\bk,\bb)}
  =
  \left\langle u_{m\bk}\middle|u_{n,\bk+\bb}\right\rangle ,
  \label{eq:wannier90-mmn-definition}
\end{equation}
in its localized-basis form
\begin{equation}
  M_{mn}^{(\bk,\bb)}
  =
  C_{m}^{\dagger}(\bk)\,
  O(\bk+\bb;\bb)\,
  C_{n}(\bk+\bb).
  \label{eq:wannier90-mmn-cp2k}
\end{equation}
Here $\bb$ is one of the neighbor vectors selected by Wannier90 and
$O(\bk+\bb;\bb)$ is obtained from real-space ``Berry'' matrices,
\begin{equation}
  O_{\mu\nu}(\bk+\bb;\bb)
  =
  \sum_{\bm{n}}
  \exp\!\left(2\pi\ii\,\bm{x}_{k+b}\cdot\bm{n}\right)
  O_{\mu\nu}^{\bm{n}}(\bb),
  \label{eq:wannier90-berry-transform}
\end{equation}
with separate cosine and sine real-space matrices in the implementation.  This
is another instance of the same image-cell transformation used for the overlap
and KS matrices, but now the operator also contains the Wannier90 neighbor
displacement.  The exported eigenvalues are gathered over k-point groups and
written to \texttt{seed.eig} in electron volts.

Reusing the symmetry-reduced SCF orbitals avoids rebuilding the same
full-mesh eigenspaces for Wannier90.  When
\texttt{REUSE\_SCF\_MOS} is active together with
\texttt{KPOINTS\_SOURCE SCF}, the code first maps every full-mesh k-point
$\bk_i$ onto an SCF representative $\bar{\bk}_{r(i)}$ and a transformation
$s_i$,
\begin{equation}
  \bk_i =
  s_i\bar{\bk}_{r(i)}
  \quad \mathrm{mod}\ \bG .
  \label{eq:wannier90-scf-map}
\end{equation}
Direct matches are reused as copies, pure time-reversal partners are obtained
by complex conjugation,
\begin{equation}
  C(-\bk)=C^{*}(\bk),
  \label{eq:wannier90-time-reversal}
\end{equation}
for scalar collinear orbitals, up to a unitary choice within degenerate
subspaces. General AO symmetry images are reconstructed as
\begin{equation}
  C_{\mu n}(s\bk)
  =
  \sum_{\nu}
  T_{\mu\nu}^{s}(\bk)\,
  \mathcal K_s[C_{\nu n}(\bk)].
  \label{eq:wannier90-ao-transform}
\end{equation}
The transformation $T^{s}$ contains the atom permutation, kind-dependent local
AO rotation matrices and image-cell phase factors, while $\mathcal K_s$ supplies
complex conjugation for antiunitary cases. CP2K
therefore reconstructs a full-mesh gauge rather than merely copying
eigenvectors from the irreducible mesh. It uses the localized Gaussian representation
used elsewhere in \Quickstep.

The reconstruction uses the lattice-periodic AO convention of
Eq.~\eqref{eq:blochsum}, in which reciprocal folding introduces no additional
atomic-position phase. The atomic images entering the overlap integrals
must also coincide with those used in the Hamiltonian. The Wannier90 interface in Section~\ref{sec:w90-projections} enforces these conventions
in the reconstruction and integral paths. Its implementation provenance
is distinguished from the mainline calculations in the Supporting
Information.

Degenerate manifolds require additional care.  The atom/AO transformation fixes
the target subspace, but the gauge within an exactly degenerate subspace is not
unique.  CP2K therefore projects the transformed vectors onto the local
$S(\bk)$ and $H(\bk)$ matrices, solves a small Ritz problem inside each
degenerate block,
\begin{equation}
  H_{\mathcal{D}}(\bk)a_j =
  \varepsilon_j
  S_{\mathcal{D}}(\bk)a_j,
  \qquad
  H_{\mathcal{D}}=C_{\mathcal{D}}^{\dagger}H(\bk)C_{\mathcal{D}},
  \quad
  S_{\mathcal{D}}=C_{\mathcal{D}}^{\dagger}S(\bk)C_{\mathcal{D}},
  \label{eq:wannier90-ritz}
\end{equation}
and exports the stabilized Ritz vectors only if the metric is nonsingular and
the residual
\begin{equation}
  \max_j
  \left\|
  H(\bk)\widetilde{C}_j
  -
  \varepsilon_j S(\bk)\widetilde{C}_j
  \right\|
  \label{eq:wannier90-ritz-residual}
\end{equation}
is below the internal tolerance.  If the requested Wannier90 band window cuts
through a degenerate SCF manifold, or if the reconstructed subspace fails the
metric/residual checks, CP2K falls back to a full-mesh diagonalization before
writing the Wannier90 files.  The diagnostic keyword
\texttt{VALIDATE\_REUSE\_SCF\_MOS} forces such a full-mesh reference to be
constructed first and compares the symmetry-reused subspaces against it.
Reuse currently requires all relevant SCF symmetry data within one k-point
process group. With several groups the exporter falls back to full-mesh
diagonalization. \texttt{PARALLEL\_GROUP\_SIZE 0} enables the one-group
layout, but is not a general performance recommendation. Finite integration
grids can also break Hamiltonian symmetry weakly. 

The \texttt{USE\_BLOCH\_PHASES} option applies CP2K's Bloch-phase gauge to
complete exported band subspaces and writes identity projections to
\texttt{seed.amn}.  It is deliberately restricted to the case where
\texttt{WANNIER\_FUNCTIONS} equals the number of exported bands. Disentangled
Wannier90 calculations still require explicit user projections.  For reused
SCF molecular orbitals, the same global gauge is applied to every
\texttt{seed.mmn} block through the phase
\begin{equation}
  \exp\!\left[
  2\pi\ii\,
  (\bm{x}_{k+b}-\bm{x}_{k})\cdot\bm{x}_{0}
  \right],
  \label{eq:wannier90-bloch-gauge}
\end{equation}
where $\bm{x}_{0}$ is the fractional center of the atomic positions.  This
keeps the exported overlaps consistent with the CP2K Bloch convention while
still allowing Wannier90 to perform the final localization. The
equal-band-count restriction applies to ordinary identity-projection export,
not to the physical-projection construction below.

\subsection{Projection, disentanglement, and Wannier interpolation}
\label{sec:w90-projections}

The Wannier90 interface connects full-mesh Gaussian-basis orbitals to physical trial projections, disentanglement, localization, and Hamiltonian interpolation. CP2K constructs eigenvalues, directed neighbor overlaps, and projection matrices in a common Bloch gauge and exchanges these quantities with Wannier90 through its file interface. The optimized rotations and real-space Hamiltonian retain explicit mappings to the original SCF band indices. 

\subsubsection{Physical projections and gauge covariance}

For a trial orbital $g_j=\sum_\mu\chi_{\mu\bk}G_{\mu j}$, the projection is
\begin{equation}
 A_{nj}(\bk)=\langle\psi_{n\bk}|g_j\rangle
 =[C^\dagger(\bk)S(\bk)G]_{nj}.
 \label{eq:w90-projection}
\end{equation}
The full AO candidate bank is obtained with $G=I$. The factor $S(\bk)$ is essential
because coefficient columns alone are not
physical projections in a nonorthogonal basis. Under a band-gauge change
$C(\bk)\mapsto C(\bk)Q(\bk)$,
\begin{equation}
 A(\bk)\mapsto Q^\dagger(\bk)A(\bk),\qquad
 M^{(\bk,\bb)}\mapsto Q^\dagger(\bk)M^{(\bk,\bb)}Q(\bk+\bb).
 \label{eq:w90-gauge-covariance}
\end{equation}
Physical projections and overlaps consequently transform together, unlike
an arbitrary identity projection that is tied to a particular eigenvector
gauge.

A selected-columns construction chooses a common trial set over the mesh with pivoted QR
factorization. It is an AO-based, selected-columns-inspired construction,
not sampling of the density matrix on a real-space grid~\cite{Damle2015}.
Every selected projection matrix is checked for rank loss at individual
k-points. Alternative anchors and column exchanges attempt a repair without
changing the band space. Hybrid trial orbitals add tetrahedral combinations
of radial $s$ and $p$ shells to the candidate bank, while retaining other
angular momenta. Numerical ties are resolved deterministically.

Automatic trial selection compares AO, hybrid, and kind-selective hybrid starts using independent Wannier90 calculations.
It also tests fixed unitary mixtures of the best original trials. Only finite,
full-rank, converged candidates can be selected. The winner has the smallest
converged total spread among the tested candidates. This finite multistart
search is not a proof of a global minimum or of unique individual centers.

\subsubsection{Disentanglement, localization, and interpolation}

For $N_b$ input bands and $N_w$ Wannier functions, a rectangular
semiunitary matrix $U^{\mathrm{dis}}(\bk)$ selects a subspace when
$N_b>N_w$, and a square unitary matrix $U(\bk)$ localizes it according to
\begin{equation}
 V(\bk)=U^{\mathrm{dis}}(\bk)U(\bk),\quad V^\dagger V=I_{N_w},\quad
 |w_{j\bR}\rangle=\frac{1}{N_k}\sum_{\bk n}
 e^{-\ii\bk\cdot\bR}|\psi_{n\bk}\rangle V_{nj}(\bk).
 \label{eq:w90-wannier}
\end{equation}
Here Bloch states are cell-normalized as above, with $U^{\mathrm{dis}}=I$ in
the isolated complete-band case. The spread functional is
\begin{equation}
 \Omega=\sum_{j=1}^{N_w}
 [\langle w_{j0}|r^2|w_{j0}\rangle
 -|\langle w_{j0}|\br|w_{j0}\rangle|^2]
 =\Omega_{\mathrm I}+\widetilde\Omega.
 \label{eq:w90-spread}
\end{equation}
Disentanglement minimizes the subspace-dependent invariant part
$\Omega_{\mathrm I}$ subject to outer and frozen energy windows.
Localization then minimizes $\widetilde\Omega$ within that
subspace~\cite{Marzari1997,Souza2001}. The outer window must contain at least
$N_w$ states at each k-point, and no more than $N_w$ states can be frozen.
The interface applies exclusions consistently to eigenvalues, overlaps,
projections, and subsequent row mappings. Both optimization stages must
converge for an automatically selected trial to be accepted.

The resulting Wannier-gauge Hamiltonian and its Fourier representation are
\begin{equation}
 H^{\mathrm W}(\bk)=V^\dagger(\bk)\varepsilon(\bk)V(\bk),\qquad
 H^{\mathrm W}(\bR)=\frac{1}{N_k}\sum_{\bk}
 e^{-\ii\bk\cdot\bR}H^{\mathrm W}(\bk).
 \label{eq:w90-hamiltonian}
\end{equation}
\texttt{WRITE\_HR} writes the native real-space Hamiltonian and
Wigner--Seitz displacement data. Interpolation must honor the associated
multiplicities and displacement convention. \texttt{WRITE\_U\_MATRICES}
writes the optimized rotations and, where applicable, $U^{\mathrm{dis}}$
and $V$. Wannier90 uses packed outer-window rows for disentangled
matrices. CP2K therefore writes \texttt{\_band\_map.dat}. Composing this
with \texttt{\_band\_indices.dat} after exclusions is necessary before
combining $V$ with the original SCF coefficients. These explicit mappings preserve the connection to the original SCF band
indices.

Collinear spin channels are treated independently, with their own band
counts, trial projections, optimized matrices, and output prefixes. This is
not spinor or spin--orbit-coupled Wannierization. The selected trial and its input data are retained for reproducible Wannier90 calculations. Losing trials cannot overwrite the winning optimized state.
These exports do not include a Wannier90 checkpoint file. Convergence
tolerances follow Wannier90's Angstrom-squared convention, whereas CP2K
reports centers and spreads in atomic units. Full-mesh reconstruction and
its fallback restrictions remain those described above.

Useful interface checks are $V^\dagger V=I$, projection rank, overlap
singular values no greater than unity for orthonormal states, reverse-neighbor
Hermitian relations, and Fourier recovery of Eq.~\eqref{eq:w90-hamiltonian}.
Centers from independent Wannier90 calculations must be compared modulo lattice translations
and Wannier permutations. Distinct local minima can have different centers
without implying an export error. Spread equality alone does not prove
identity of the localized orbitals.

\subsection{Band structures, densities of states, and population analysis}

The Bloch matrices also support band-resolved analysis at a fixed SCF
potential. Band-structure calculations use \texttt{GENERAL} k-point paths
through specified special points, diagonalize the KS problem along the path,
and report k-resolved energies and occupations. Density-of-states (DOS)
calculations instead require Brillouin-zone quadrature and can use an
independent Monkhorst--Pack mesh, with a single point in nonperiodic
directions. A band path is therefore not a replacement for an integration
mesh, and convergence of the SCF density does not by itself establish
convergence of a DOS evaluated on a separate mesh.

Population analysis additionally retains the nonorthogonal AO metric.
The L\"owdin population associated with basis function $\mu$ is
\begin{equation}
  \rho_{\mu}^{\mathrm{L\ddot{o}wdin}}
  =\textbf{1}
  \sum_{\bk} w_{\bk}
  \left[
  S^{1/2}(\bk)P(\bk)S^{1/2}(\bk)
  \right]_{\mu\mu},
  \label{eq:lowdin}
\end{equation}
where $P(\bk)$ includes the orbital occupations and $S^{1/2}(\bk)$ is the
Hermitian positive square root of the overlap matrix\cite{Hehn2022, Mayhall2010}. The complex matrix
algebra is retained throughout. Summing over all basis functions recovers
$\sum_{\bk}w_{\bk}\Tr[S(\bk)P(\bk)]$, the electron count in the same spin
convention. Atomic populations are obtained by grouping the corresponding
basis functions. Individual atomic populations are basis dependent.
On a symmetry-reduced mesh, atom-resolved analysis requires full-orbit
matrix reconstruction or equivalent symmetry averaging of the orbital
contributions. Scalar representative weights alone do not generally
recover the populations of individual atoms.

\subsection{Supporting TREXIO and CASINO exports}

Orbital exchange with other electronic-structure methods requires the same
atomic-image and Bloch-gauge information, even when no inter-k-point
neighbor graph is transferred. TREXIO~\cite{Posenitskiy2023} provides a
general archival and exchange export. For k-point calculations it writes the
periodic cell, k-point coordinates and weights, k-resolved MO energies,
occupations, spin labels, real and imaginary MO coefficients, and an
MO-to-k-point index.  The TREXIO writer rephases complex k-point MO coefficients so that the exported
\texttt{nucleus\_coord} values and the Bloch sums assumed by external TREXIO
consumers are in the same atomic gauge. If atom $I$ has fractional coordinate
$\bm{s}_I$, let $\bm{s}_I^{\mathrm{wrap}}$ be its image selected by the
actual periodic-boundary routine. For every AO $\mu$ on atom $I$, the exported
coefficient is therefore adjusted as
\begin{equation}
  C_{\mu n}^{\mathrm{TREXIO}}(\bk)
  =
  \exp\!\left[-2\pi\ii\,\bm x_k\cdot\bm{g}_{I}\right]\,
  C_{\mu n}^{\mathrm{CP2K}}(\bk),
  \qquad
  \bm{g}_{I}=\mathrm{NINT}(\bm{s}_I^{\mathrm{wrap}}-\bm{s}_I),
  \label{eq:trexio-atom-gauge}
\end{equation}
with the real and imaginary parts rotated consistently before the corresponding
TREXIO MO-coefficient arrays are written. This formulation respects the
image convention at half-cell boundaries, where an assumed universal
$-\mathrm{NINT}(\bm s_I)$ can be inappropriate.

The current writer supports \texttt{FULL\_KPOINT\_GRID}. For an
atomic-symmetry-reduced SCF object it prepares the full regular mesh using
the symmetry-reconstruction machinery also used by Wannier90, with a
diagonalization fallback. The inspected guard enters this path only when
\texttt{SYMMETRY} is enabled and \texttt{FULL\_GRID} is disabled. It must
not be interpreted as unconditional expansion of every time-reversal-reduced
object with \texttt{SYMMETRY F}. Exported weights and MO-to-k-point indices
belong to the selected export object. The archive records orbitals and their k-point metadata. Wannier90's
neighbor graph, inter-point overlaps, projection selection, and localization
remain separate operations.

CP2K also provides a native CASINO \texttt{gwfn.data} writer for quantum
Monte Carlo applications~\cite{Needs2020}. It reorders and normalizes the
Gaussian spherical harmonics, applies the atom-image gauge, and orders real
boundary points and complex conjugate k-point pairs according to the target
format. The current path requires a complete MO basis, supports spherical
shells through $g$ and up to two collinear spin channels, and can reconstruct
an atomic-symmetry-reduced regular mesh through the shared preparation path.
Exporting orbitals does not itself perform or validate a many-body calculation.
For twist averaging, the many-electron boundary condition is
\begin{equation}
 \Psi(\ldots,\br_i+\bR,\ldots)=e^{\ii\bm\theta\cdot\bR}
 \Psi(\ldots,\br_i,\ldots),\qquad
 \langle E\rangle_{\theta}\simeq\sum_t v_t E(\bm\theta_t).
\end{equation}
The relation between a primitive-cell orbital mesh, a supercell, and its
many-electron twist must therefore be chosen explicitly by the downstream
calculation~\cite{Lin2001}. A DFT integration weight is not automatically a
validated quantum Monte Carlo sampling prescription.

\subsection{Response and transport}

Time-dependent density-functional perturbation theory (TDDFPT)
is connected in the independent-particle limit. K-point calculations
require complex wavefunctions and \texttt{KERNEL NONE}. Full kernels,
smearing, spin flip, orbital-energy corrections, exciton descriptors,
natural-transition-orbital analysis, and k-point TDDFPT forces are not yet implemented.
Transport paths are also present.  The SMEAGOL
interface explicitly requires a k-point calculation for its
electronic-transport run type, while the internal nonequilibrium
Green's-function (NEGF) machinery has
k-resolved contact and transport input paths but still keeps some device
force-evaluation combinations guarded.

\section{Hybrid exchange and many-body methods}
\label{sec:advanced-kpoint-methods}

Hybrid functionals and many-body perturbation theory extend the momentum
structure beyond a single KS wavevector. They combine pairs of electronic
momenta, auxiliary-basis matrices, and, for correlated methods, screened
Coulomb quantities.  They are therefore treated separately from the
single-particle post-SCF interfaces above.  The important design point is the
same as in the KS part of the code. The implementation does not introduce a
new plane-wave representation for these methods.  It reuses image-cell
Gaussian and auxiliary-basis matrices, transforms only the objects that carry a
Bloch phase, and returns the resulting matrices to the localized-basis data
model.

\subsection{Hybrid exchange and auxiliary-density-matrix methods}

The production route for hybrid-functional k-point calculations in
\CPtwoK/\Quickstep{} is resolution-of-identity Hartree--Fock exchange
(RI-HFX)~\cite{Bussy2024}. With cell-normalized Bloch orbitals, the
exchange energy per cell and spin channel can be written as
\begin{equation}
  E_{x}^{\mathrm{HF},\sigma}
  =
  -\frac{1}{2}
  \sum_{\bk\bk'} w_{\bk}w_{\bk'}
  \sum_{ij} f_{i\bk}^{\sigma} f_{j\bk'}^{\sigma}
  \int_{\Omega_{\mathrm{cell}}}\dd\br
  \int_{\mathbb R^3}\dd\br'\,
  \psi_{i\bk}^{\sigma *}(\br)
  \psi_{j\bk'}^{\sigma}(\br)
  v(\br-\br')
  \psi_{j\bk'}^{\sigma *}(\br')
  \psi_{i\bk}^{\sigma}(\br') .
  \label{eq:hfx-energy-kpoints}
\end{equation}
Here $v$ is the selected exchange kernel, including its periodic or
truncated-interaction prescription. For the unscreened Coulomb interaction,
$v(\br-\br')=1/|\br-\br'|$ in atomic units. The home-cell integral
fixes the per-cell normalization. Occupations are resolved into individual spin channels with
$0\leq f^\sigma\leq1$. A restricted calculation's double occupancy must
first be split between the two channels in this expression.
In a localized Gaussian implementation, Eq.~\eqref{eq:hfx-energy-kpoints}
becomes a contraction over density matrices, three-center overlap/Coulomb
objects, and a k-dependent auxiliary Coulomb metric.  The two-center metric is
obtained by the same lattice-sum convention as the one-electron matrices,
\begin{equation}
  V_{PQ}(\bq) =
  \sum_{\bm{n}}
  \exp\!\left(2\pi\ii\,\bm{x}_{q}\cdot\bm{n}\right)
  (P\,0|Q\,\bm{n}),
  \label{eq:coulomb-2c}
\end{equation}
where $(P\,0|Q\,\bm{n})$ denotes a two-center Coulomb integral between an
auxiliary basis function in the home cell and one in image cell $\bm{n}$.
The momentum transfer $\bq$ is generated from pairs of electronic k-points.
Schematically, the RI-HFX contribution to the KS matrix has the form
\begin{equation}
  K_{\mu\nu}^{\sigma}(\bk)
  \approx
  \sum_{\bk'} w_{\bk'}
  \sum_{\lambda\kappa}
  P_{\lambda\kappa}^{\sigma}(\bk')
  \sum_{PQ}
  B_{\mu\lambda}^{P}(\bk,\bk')
  \left[V^{-1}(\bk-\bk')\right]_{PQ}
  B_{\nu\kappa}^{Q *}(\bk,\bk'),
  \label{eq:ri-hfx-schematic}
\end{equation}
where $B_{\mu\lambda}^{P}$ denotes the transformed three-center quantity.
Here $K$ is defined as the positive exchange contraction. The hybrid
Hamiltonian contains $-\alpha K$.
Equation~\eqref{eq:ri-hfx-schematic} is schematic, but it shows the essential
software requirement. The k-point code must provide consistent phases for
orbital-basis matrices and auxiliary-basis Coulomb matrices at the same
momentum transfer.

The auxiliary density matrix method (ADMM)~\cite{Guidon2010} reduces the cost of hybrid exchange
by projecting the density matrix from the primary orbital basis to a smaller
auxiliary orbital basis.  In its usual hybrid form the correction can be
written schematically as
\begin{equation}
  E_{\mathrm{xc}}^{\mathrm{ADMM}}[P]
  =
  E_{\mathrm{xc}}^{\mathrm{semi-local}}[P]
  +
  \alpha
  \left(
  E_{x}^{\mathrm{HF}}[\widetilde{P}]
  -
  E_{x}^{\mathrm{semi-local}}[\widetilde{P}]
  \right),
  \label{eq:admm-schematic}
\end{equation}
where $\widetilde{P}$ is the projected density matrix and $\alpha$ is the
hybrid mixing coefficient.  With k-points, the projection and back-projection
must be applied to the real and imaginary density-matrix components for each
representative k-point and must remain compatible with the symmetry expansion
described in Section~\ref{sec:k290-spglib}.  This is why ADMM is not merely a
basis-set option in the present implementation, but another consumer of the
k-point matrix-transformation layer.

\subsection{Many-body perturbation theory}

The random-phase approximation (RPA) and GW approximation, which uses
the one-electron Green's function $G$ and screened interaction
$W$~\cite{Wilhelm2016,Wilhelm2017,Pasquier2025}, use the same phase conventions for
auxiliary-basis matrices, but they also introduce transfer momenta.  The
noninteracting response in an auxiliary basis may be written as
\begin{equation}
  \chi_{PQ}^{0}(\bq,\ii\omega)
  =
  \sum_{\bk}
  w_{\bk}
  \sum_{nm\sigma}
  \frac{
  f_{n\bk}^{\sigma}
  -
  f_{m,\bk+\bq}^{\sigma}}
  {\ii\omega
  +
  \varepsilon_{n\bk}^{\sigma}
  -
  \varepsilon_{m,\bk+\bq}^{\sigma}}
  M_{nm}^{P,\sigma}(\bk,\bq)
  M_{nm}^{Q,\sigma *}(\bk,\bq),
  \label{eq:chi0-kpoints}
\end{equation}
where $M_{nm}^{P,\sigma}(\bk,\bq)$ is the auxiliary-basis transition density
between a state at $\bk$ and a state at $\bk+\bq$.  The screened interaction is
then built from the Coulomb metric in Eq.~\eqref{eq:coulomb-2c},
\begin{equation}
  W(\bq,\omega)
  =
  \left[
  I - V(\bq)\chi^{0}(\bq,\omega)
  \right]^{-1}
  V(\bq),
  \label{eq:screened-interaction}
\end{equation}
and the corresponding RPA correlation energy has the standard trace form
\begin{equation}
  E_{c}^{\mathrm{RPA}}
  =
  \frac{1}{2\pi}
  \int_{0}^{\infty}
  \sum_{\bq} w_{\bq}
  \operatorname{Tr}
  \left[
  \ln\!\left(I-V(\bq)\chi^{0}(\bq,\ii\omega)\right)
  +
  V(\bq)\chi^{0}(\bq,\ii\omega)
  \right]\dd\omega .
  \label{eq:rpa-energy-kpoints}
\end{equation}
For GW, the self-energy of a quasiparticle state can be expressed
schematically as
\begin{equation}
  \Sigma_{n\bk}(\omega)
  =
  \frac{\ii}{2\pi}
  \sum_{\bq} w_{\bq}
  \sum_m
  \int
  M_{nm}^{P}(\bk,\bq)
  G_{m,\bk+\bq}(\omega+\omega')
  W_{PQ}(\bq,\omega')
  M_{nm}^{Q *}(\bk,\bq)
  \,\dd\omega',
  \label{eq:gw-self-energy-kpoints}
\end{equation}
with implicit summation over auxiliary-basis indices $P$ and $Q$ and the
usual frequency-contour prescription. The transfer-momentum sign is chosen
consistently with the $\bk\to\bk+\bq$ transition density above.

The implementation includes cubic-scaling RPA and $G_0W_0$ self-energy
paths, with separate sampling of screening and self-energy momenta.
The accompanying source inventory identifies regression test inputs for
spin--orbit gaps, local densities of states, and band dispersions.  The implementation groups lattice vectors
into ranges tied to the Monkhorst--Pack mesh and respects nonperiodic
directions by enforcing one k-point and zero translation range along those
axes.  Some combinations remain intentionally guarded, including singular-value
decomposition integrals with k-points and full low-scaling RPA k-points
combined with GW. These guards are important because, for correlated methods, a wrong
k-point phase or an inconsistent transfer-momentum grid can produce plausible
but incorrect spectra. In particular, periodic GW sampling belongs to its
own correlated-method infrastructure. Enabling \texttt{DFT/KPOINTS} does
not automatically enable every GW k-point route. Automatic auxiliary-basis generation, common imaginary-time propagators,
and consistent band-window mappings support these calculations.
Auxiliary bases, frequency grids, empty states, and transfer momenta
nevertheless remain independent convergence parameters.

\section{Numerical verification}
\label{sec:verification}

The numerical tests separately examine convergence of the physical
quadrature and equivalence of different representations of the same mesh,
using semiconducting silicon and metallic aluminium. These calculations
use the version-identified mainline implementation and are complemented
by algebraic unit tests. The Supporting Information
lists the broader regression coverage and distinguishes source-level
support from tests executed for this paper.

\subsection{Calculation settings}

Eighteen converged calculations were performed with the unmodified mainline
source documented in the accompanying calculation records. Conventional cubic cells containing
eight Si atoms ($a=5.43$~\AA) and four Al atoms ($a=4.05$~\AA) provide
semiconducting and metallic examples, respectively. We use the
Perdew--Burke--Ernzerhof (PBE) functional~\cite{Perdew1996},
DZVP-MOLOPT-SR-GTH Gaussian bases~\cite{VandeVondele2007},
and GTH-PBE pseudopotentials with four and three valence electrons for
Si and Al~\cite{Goedecker1996,Krack2005}. The density and relative grid
cutoffs are 600 and 60~Ry, the SCF threshold is $10^{-9}$, and standard
diagonalization with Broyden mixing is used. Al employs Fermi--Dirac
occupations at 1000~K. Its reported energy includes the electronic entropy
term and is therefore the electronic free energy. All cases use explicit
complex wavefunctions, one k-point group, one OpenMP thread and one Basic Linear Algebra
Subprograms (BLAS) thread. The executable was built with GNU Fortran 16.1, DBCSR, OpenBLAS,
FFTW, Libint, Libxc and SPGLIB. Complete inputs,
outputs and build metadata accompany the paper.

\subsection{Quadrature convergence}

Table~\ref{tab:mesh-validation} separates quadrature convergence from
implementation equivalence. Between the full $6^3$ and $8^3$ meshes, the Si
energy changes by $0.043$~meV per atom, whereas the Al free energy still changes
by $10.721$~meV per atom. Thus even with finite-temperature occupations,
a metallic example requires a substantially denser mesh than this
semiconducting example. The Al chemical potential is nonmonotonic over
this sequence. Neither agreement between two representations of a mesh
nor a small SCF residual establishes convergence with respect to that mesh.
Quadrature convergence must also be distinguished from Gaussian-basis and
pseudopotential errors. The UZH protocol~\cite{Mirhosseini2026UZH} separates
these latter contributions through matched-pseudopotential plane-wave and
all-electron reference calculations, providing a complementary assessment
of the underlying basis and pseudopotential approximations.

\subsection{Representation equivalence}

Table~\ref{tab:symmetry-validation} compares identical $4^3$ quadratures.
Time reversal reduces 64 solved points to 32, while either K290 or SPGLIB
reduces them to four. This sixteenfold reduction in the number of
eigenproblems is an algorithmic count, not a measured sixteenfold
wall-time speedup. The largest absolute cell-energy/free-energy difference
among these representations is $1.4\times10^{-12}\,E_h$. Force-component
differences remain below $2.1\times10^{-15}\,E_h/a_0$.
The Al stress difference under atomic-symmetry reduction is
$0.098$~bar, appreciably larger than its roundoff-level energy difference.
Reporting stress as well as energy therefore provides a distinct check.
These high-symmetry cells have essentially vanishing physical forces,
so they do not replace finite-difference force tests on distorted structures.

\input{validation_tables.tex}

Requesting \texttt{LATTICE\_FFT ON} instead of \texttt{OFF} changes the
cell energies by less than $2.0\times10^{-13}\,E_h$ in these cases.
Because a guarded implementation can fall back for individual
transformations, this comparison alone does not prove that every
transformation used the FFT path. Three additional upstream unit tests
were executed successfully, namely \texttt{kpoint\_lattice\_fft\_unittest},
\texttt{kpsym\_unittest}, and \texttt{kpoint\_smearing\_unittest}.
They separately exercise the transform algebra, symmetry-orbit guards,
and occupations. The symmetry test intentionally checks a warning-producing
fallback case.

The examples establish fixed-mesh equivalence for the tested cells and
execution layout. They do not measure parallel speedup or nonzero-force
accuracy. The Supporting Information records the execution configuration,
additional diagnostics, and method-specific compatibility conditions.
In particular, symmetry reduction must follow the symmetry of the
Hamiltonian actually solved, while response and correlated methods require
their own sampling and convergence checks.

\section{Conclusions}

Symmetry-reduced k-point sampling in a localized nonorthogonal basis is a
matrix-reconstruction problem as well as a quadrature problem. The
formulation presented here connects sparse image-cell operators,
irreducible-point solutions, and full symmetry orbits through explicit
atomic permutations, orbital rotations, and lattice phases. Retaining
these transformations preserves the basis metric and allows scalar,
matrix, and wavefunction quantities to use a common reciprocal-space
description. K290 and SPGLIB supply the symmetry information, while the
implementation in \CPtwoK{} retains the real-space Gaussian representation
and the existing density and potential machinery.

The numerical examples establish fixed-quadrature equivalence and the
separate need for mesh convergence. For Si and Al, symmetry reduction from 64 points to four
representatives preserves the cell energy to within
$1.5\times10^{-12}$ hartree at fixed quadrature. The mesh series separately
demonstrates the different convergence requirements of an insulator and a
metal. These comparisons concern the reconstruction of integrated
observables; they do not substitute for validation of downstream
property calculations.

The Wannier90 interface connects the same
orbital representation to projection, disentanglement, localization, and
Hamiltonian interpolation. The shared requirement is a
consistent gauge and inter-point connectivity, which cannot be recovered
from integration weights or band energies alone. Hybrid and many-body
applications add auxiliary-basis and transfer-momentum transformations
with their own sampling requirements.

The resulting methodological picture extends from symmetry-reduced
Brillouin-zone integration to gauge-consistent analysis of periodic
electronic states. Its implementation and reproducible examples provide
a concrete reference for connecting these tasks in localized-orbital
electronic-structure methods.

\section*{Code Availability}

\CPtwoK{} is open-source software distributed under the GNU General
Public License, version 2 or later, at
\url{https://github.com/cp2k/cp2k}. The mainline calculations and Wannier90 interface correspond to the separate implementations
described in this work. Their exact source
provenance and validation scope are recorded in the accompanying technical
archive. Input generators, analysis scripts, source-inspection records,
and build metadata accompany the numerical data.

\section*{Data Availability}

The numerical validation package is maintained at
\url{https://github.com/DCM-Uni-Paderborn/cp2k-kpoint-data}. It contains
input files, raw calculation outputs, numerical tables, analysis
scripts, and build metadata. The numerical values underlying both tables are retained at full precision, together with the regeneration script and checksums.
The Supporting Information describes the calculations and the separate
source versions.

\section*{CRediT Authorship Contribution Statement}

Johann V. Pototschnig: Software, Investigation, Validation, Writing -- review and editing. 
Jan Kloppenburg: Software, Investigation, Writing --
review and editing. 
Shaojie He: Software.
Chengze Zhou: Software. 
Shiyang Wang: Software, Writing -- review and editing.
Sam Azadi: Validation, Writing -- review and editing.
Hossein Mirhosseini: Methodology, Validation, Writing -- review and editing. 
J\"urg Hutter: Methodology, Software, Writing --
review and editing. 
Thomas D. K\"uhne: Conceptualization, Methodology,
Supervision, Project administration, Writing -- original draft, Writing --
review and editing.
% Proposed roles require approval by every author; see submission_checklist.md.

\section*{Declaration of Competing Interest}

The authors declare no known competing financial interests or personal
relationships that could have appeared to influence the work reported in this
paper.
% Draft declaration: obtain all-author confirmation before submission.

% Funding and computing allocations await the authors' confirmation.
% Add the approved statement before submission; see submission_checklist.md.

\section*{Declaration of Generative AI and AI-Assisted Technologies in the
Manuscript Preparation Process}

During the preparation of this work the authors used Terok~\cite{Vyskocil2026Terok} with the GPT-5.6 Sol model (OpenAI) to
support language editing, mathematical
exposition, data-analysis and plotting scripts,
consistency checks, and LaTeX preparation.
After using this tool, the authors reviewed and edited the
content as needed and take full responsibility for the content of the published
article.
% The stated author review and responsibility must be confirmed before submission.

\section*{Acknowledgements}

The authors thank the \CPtwoK{} developer community for the continued
development of \Quickstep{} and its k-point infrastructure, and the
developers of SPGLIB, Wannier90, TREXIO, and CASINO for their open software.
We thank Zilong He for reproducible reports on MPI-distributed k-point
orbital output.
% Add project-specific computing acknowledgements only after author confirmation.

\end{document}

%% file: validation_tables.tex
\begin{table}[htbp]
\centering\small\setlength{\tabcolsep}{4pt}
\caption{Mesh dependence for conventional cubic Si and Al cells. The tabulated quantity is energy per atom for Si and electronic free energy per atom for Al at 1000 K. Differences refer to the corresponding full $8^3$ mesh. This mesh sequence does not establish the infinite-mesh or complete-basis limit. Chemical potentials use CP2K's internal energy zero.}
\label{tab:mesh-validation}
\begin{tabular}{llrrr}
\toprule
System & Mesh & $E/N$ or $\mathcal F/N$ ($E_h$) & $\Delta$ (meV/atom) & $\mu$ ($E_h$) \\
\midrule
Si & $1^3$ & -3.8894918076 & 1277.879 & -- \\
Si & $2^3$ & -3.9334572345 & 81.519 & -- \\
Si & $4^3$ & -3.9364036379 & 1.344 & -- \\
Si & $6^3$ & -3.9364514258 & 0.043 & -- \\
Si & $8^3$ & -3.9364530128 & 0.000 & -- \\
Al & $1^3$ & -2.0101729640 & 1676.252 & 0.20014793 \\
Al & $2^3$ & -2.0660631977 & 155.401 & 0.23145703 \\
Al & $4^3$ & -2.0701747224 & 43.521 & 0.26198590 \\
Al & $6^3$ & -2.0713800877 & 10.721 & 0.25412969 \\
Al & $8^3$ & -2.0717740941 & 0.000 & 0.25645631 \\
\bottomrule
\end{tabular}
\end{table}
\begin{table}[htbp]
\centering\small\setlength{\tabcolsep}{4pt}
\caption{Same-mesh comparisons at $4^3$, relative to the direct full-grid result. Columns report solved k-points, absolute cell-energy/free-energy differences, maximum force-component differences, and maximum stress-component differences. All runs use one OpenMP thread. The FFT label records the requested option. Independent transform unit tests also check the FFT algebra.}
\label{tab:symmetry-validation}
\begin{tabular}{llrrrr}
\toprule
System & Mode & $N_k$ & $|\Delta E|$ ($E_h$) & $\|\Delta F\|_\infty$ ($E_h/a_0$) & $\|\Delta\sigma\|_\infty$ (bar) \\
\midrule
Si & Full & 64 & $0$ & $0$ & $0$ \\
Si & Time reversal & 32 & $1.1\times10^{-14}$ & $9.6\times10^{-16}$ & $2.4\times10^{-10}$ \\
Si & K290 & 4 & $1.4\times10^{-14}$ & $2.0\times10^{-15}$ & $2.3\times10^{-10}$ \\
Si & SPGLIB & 4 & $1.4\times10^{-14}$ & $2.0\times10^{-15}$ & $2.3\times10^{-10}$ \\
Si & Full, FFT requested & 64 & $7.1\times10^{-15}$ & $1.9\times10^{-15}$ & $4.3\times10^{-10}$ \\
Al & Full & 64 & $0$ & $0$ & $0$ \\
Al & Time reversal & 32 & $1.4\times10^{-12}$ & $1.1\times10^{-15}$ & $1.1\times10^{-06}$ \\
Al & K290 & 4 & $5.3\times10^{-13}$ & $9.3\times10^{-16}$ & $9.8\times10^{-02}$ \\
Al & SPGLIB & 4 & $5.3\times10^{-13}$ & $9.3\times10^{-16}$ & $9.8\times10^{-02}$ \\
Al & Full, FFT requested & 64 & $1.9\times10^{-13}$ & $7.3\times10^{-16}$ & $7.0\times10^{-08}$ \\
\bottomrule
\end{tabular}
\end{table}